\documentstyle[11pt]{article}\textheight 230mm\textwidth 150mm
            \newcommand{\be}{\begin{eqnarray}}
            \newcommand{\ee}{\end{eqnarray}}
            \newcommand{\eel}[1]{\label{#1}\end{eqnarray}}
\newcommand{\e}[1]{\label{e:#1}\end{eqnarray}}
     \newcommand{\eg}{{\em e.g.\ }}

 \newcommand{\Ga}{{\Gamma}}
            \newcommand{\la}{{\lambda}}
            
            \newcommand{\del}{{\delta}}
           \newcommand{\ra}{{\rightarrow}}
 \newcommand{\lea}{{\leftarrow}}
            \newcommand{\lra}{{\leftrightarrow}}

            \newcommand{\beq}{\begin{quote}}
            \newcommand{\eq}{\end{quote}}
            \newcommand{\Om}{\Omega}
            \newcommand{\al}{\alpha}
            \newcommand{\halv}{\frac{1}{2}}
            \newcommand{\ben}{\begin{enumerate}}
            \newcommand{\een}{\end{enumerate}}
            \newcommand{\bit}{\begin{itemize}}
            \newcommand{\ei}{\end{itemize}}
\newcommand{\ve}{{\varepsilon}}
    		
            \newcommand{\nn}{\nonumber}
            \newcommand{\r}[1]{(\ref{e:#1})}
            \newcommand{\edfl}[1]{\label{#1}\end{df}}
            
\begin{document}
            \begin{titlepage}
            \newpage
            \noindent
            G\"oteborg ITP 94-30\\
            January 1995\\
												hep-th/9502030
            \vspace*{30 mm}
            \begin{center}{\LARGE\bf Completely anticanonical form of
 Sp(2)-symmetric
Lagrangian quantization}\end{center}
            \vspace*{15 mm}
        \begin{center}{\large \bf Igor Batalin}
\footnote{On leave of absence from
P.N.Lebedev Physical Institute, 117924  Moscow, Russia } and
{\large \bf Robert Marnelius} \\
          \vspace*{10 mm} {\sl
            Institute of Theoretical Physics\\ Chalmers University of
            Technology\\ S-412 96  G\"{o}teborg, Sweden}\end{center}
            \vspace*{27 mm}
            \begin{abstract}
            The Sp(2)-symmetric Lagrangian quantization scheme is
represented in a
completely anticanonical form. Antifields are assigned to all field
variables
including former "parametric" ones $\pi^{Aa}$. The antibrackets $(F, G)^a$
as well as
the operators $\triangle^a$ and $V^a$ are extended to include the new
anticanonical
pairs $\pi^{Aa}$, $\bar{\phi}_A$. A new version of the gauge fixing mechanism
in the Lagrangian
effective action  is proposed. The corresponding functional integral is shown
to be gauge
independent. \end{abstract} \end{titlepage}
            \newpage
            \setcounter{page}{1}
            \section{Introduction}
In papers \cite{BLT1,BLT2,BLT3} an Sp(2)-symmetric scheme of the Lagrangian
 (field-antifield)
quantization was developed. (The global Sp(2) symmetry
allows one to consider ghost
and antighost field variables in a uniform way.)
Geometric aspects as well as various possibilities of interpretation of this
formalism have been
considered in papers \cite{MH,BCG,GH,DD,LT}.

In the Sp(2)-symmetric scheme one deals with a complete set of field
variables $\phi^A$, which
includes the original fields, the ghosts, the antighosts and the Lagrange
multipliers. In addition
one assigns to each $\phi^A$ three kinds of antifields: $\phi^*_{A1}$,
$\phi^*_{A2}$ and
$\bar{\phi}_A$. From the Hamiltonian point of view $\phi^*_{A1}$ and
$\phi^*_{A2}$ correspond to the
sources of BRST and antiBRST field transformations, $[\phi^A, \Om^1]$
and $[\phi^A, \Om^2]$, while
$\bar{\phi}_A$ corresponds to the source of their combined transformation,
$\varepsilon_{ab}[[\phi^A,
\Om^a], \Om^b]$.

Unfortunately,  these three kinds of antifields enter the present Sp(2)
symmetric formalism
in a nonsymmetric way. While $\phi^*_{Aa}$ are anticanonically conjugate
to $\phi^A$ in the usual sense,
$\bar{\phi}_A$ do not possess their own conjugate fields from the very
beginning. On the other
hand, when a gauge fixing procedure was introduced into the Sp(2)
symmetric theory one had to make
use of some auxiliary field variables $\pi^{Aa}$ to parametrize the
differential operator containing
the gauge fixing function \cite{BLT1,BLT2,BLT3}.

The main idea of the present paper is to identify the previous
parametric variable $\pi^{Aa}$ with
an auxiliary  field variable which is conjugated to the antifield
$\bar{\phi}_A$ in the usual sense.
Then we define an extended antibracket $(F, G)^a$ and an operator
$\triangle^a$ into which all
fields-antifields enter on equal footing as anticanonical pairs. An
extended version of the master
equation whose solution generally depends on the complete set of
anticanonical pairs is then set up.
  In this way we are  able to generalize the description of abelian Lagrangian
surface to cover the Sp(2)-symmetric case. We formulate a completely
anticanonical version of
generalized BRST-antiBRST transformations and claim that the functional
integral  is
gauge independent.    This is explicitly verified for a simple consistent
 set of gauge fixing conditions which are explicitly solved with
respect to $\phi^*_{Aa}$.

        \section{Main definitions}

Let $\phi^A$, $\ve(\phi^A)\equiv\ve_A$, be a complete set of field
variables including original
fields, ghosts, antighosts and Lagrange multipliers. Let us assign to
each of them a pair of
antifields $\phi^*_{Aa}$ ($a=1,2$), $\ve(\phi^*_{Aa})\equiv\ve_A+1$. Next
let us introduce a set of
pairs of auxiliary field variables $\pi^{Aa}$, $\ve(\pi^{Aa})\equiv\ve_A+1$,
whose antifields are
$\bar{\phi}_A$,  $\ve(\bar{\phi}_A)\equiv\ve_A$. In terms of these variables
we define an extended
antibracket by
\be
&&(F, G)^a\equiv F\left(
\frac{\stackrel{\lea}{\partial}}{\partial\phi^A}
\frac{\stackrel{\ra}{\partial}}{\partial\phi^*_{Aa}}+\ve^{ab}
\frac{\stackrel{\lea}{\partial}}{\partial\pi^{Ab}}
\frac{\stackrel{\ra}{\partial}}{\partial\bar{\phi}_A}
\right)G-(F\lra G)(-1)^{(\ve(F)+1)(\ve(G)+1)}
\e{1}
and the corresponding operator $\triangle^a$ given by
\be
&&\triangle^a\equiv(-1)^{\ve_A}\frac{\stackrel{\ra}
{\partial}}{\partial\phi^A}
\frac{\stackrel{\ra}{\partial}}{\partial\phi^*_{Aa}}+\ve^{ab}(-1)^{\ve_A+1}
\frac{\stackrel{\ra}{\partial}}{\partial\pi^{Aa}}
\frac{\stackrel{\ra}{\partial}}{\partial\bar{\phi}_A}
\e{2}
The antibrackets \r{1} satisfy the generalized Jacobi identities
\be
&&((F, G)^{\{a}, H)^{b\}}(-1)^{(\ve(F)+1)(\ve(H)+1)}
+\mbox{cycle}(F,G,H)\equiv 0
\e{3}
and $\triangle^a$ satisfies
\be
&&\triangle^{\{a}\triangle^{b\}}=0
\e{4}
\be
&&\triangle^{\{a}(F, G)^{b\}}=(   \triangle^{\{a} F, G)^{b\}}
-(F,\triangle^{\{a}
G)^{b\}}(-1)^{\ve(F)}
\e{5}
\be
&&\triangle^a(FG)= (  \triangle^{a }F) G+(F, G)^a(-1)^{\ve(F) }+
F (\triangle^{a }G)(-1)^{\ve(F) }
\e{6}
The curly bracket denotes symmetrization of $a$ and $b$. Notice that
formula \r{6} may be
considered to be an alternative definition of the antibracket \r{1}.
Let us also introduce the
extended operator $\bar{\triangle}^a$:
\be
&&\bar{\triangle}^a\equiv\triangle^a+\frac{i}{\hbar}V^a
\e{7}
where
\be
&&V^a\equiv\halv\left(\ve^{ab}\phi^*_{Ab}
\frac{\partial}{\partial\bar{\phi}_A}
-\pi^{Aa}(-1)^{\ve_A}\frac{\partial}{\partial\phi^A}\right)
\e{8}
One may easily check that the following properties hold
\be
&&V^a(F, G)^b=(V^aF, G)^b-(-1)^{\ve(F)}(F, V^aG)^b
\e{9}
\be
&&V^{\{a}V^{b\}}=0
\e{10}
\be
&&\triangle^{a} V^{b}+V^{b}\triangle^{a}=0
\e{11}
{}From \r{10} and \r{11} it follows then that
\be
&&\bar{\triangle}^{\{a}\bar{\triangle}^{b\}}=0
\e{12}
Notice that our definition of $V^a$, \r{8}, differs from the one given in
\cite{BLT1,BLT2,BLT3}. As
a consequence  our formulas \r{9} and \r{11} are valid
 without symmetrization in the
indices $a$ and $b$.

            \section{Gauge independent functional integral representation}

The quantum action $W(\phi, \phi^*, \pi, \bar{\phi};\hbar)$
is defined to be a solution
of the following quantum master equation
\be
&&\bar{\triangle}^a\exp\{\frac{i}{\hbar}W\}=0
\e{13}
or equivalently
\be
&&\halv(W, W)^a+V^aW=i\hbar{\triangle}^aW
\e{14}
We define then the field-antifield functional integral to be
\be
&&Z=\int [d\phi] [d\phi^*] [d\pi] [d\bar{\phi}] [d\la] \exp\{\frac{i}{\hbar}
[W+X]\} \e{15}
where $X(\phi, \phi^*, \pi, \bar{\phi};\hbar)$ is a hypergauge fixing action
depending on a new variable
$\la^A$, $\ve(\la^A)=\ve_A$, corresponding to the
 hypergauge invariance of $W$. (If
$X$ is linear in $\la^A$ they  are Lagrange multipliers.)
 $X$ is required to satisfy
the following quantum master equation
\be
&&\halv (X,X)^a-V^aX=i\hbar\Delta^a X
\e{16}
which is the same equation as \r{14} apart from the
opposite sign of the $V$-term. We
expect the classical part of $X$ to have the structure
\be
&&\left.X\right|_{\hbar=0}=G_A\la^A+{\cal K}Y
 \e{17}
where $Y$  and $G_A$  are functions while
${\cal K}$ is the differential
operator
\be
&&{\cal K}\equiv \ve_{ab}V^aV^b
\e{18}
One may notice   that $Y$ is only determined up to functions $A$ satisfying
\be
&&{\cal K}A=0
\e{19}
$A$ may \eg have the form $A=V^aR$ any $R$. When \r{17} is inserted into
\r{16} we find
that the functions $Y$  and $G_A$ are subjected to the following conditions
\be
&&(G_A, G_B)^a=0
\e{192}
\be
&&({\cal K}Y, G_A)^a=V^aG_A
\e{193}
\be
&&({\cal K}Y, {\cal K}Y)^a=0
\e{194}

Let $\Ga$ denote the complete set of field-antifield variables.
We assert then that \r{15} is
invariant under the  following transformation:
\be
&&\del\Ga\equiv(\Ga, -W+X)^a\mu_a-2V^a\Ga(-1)^{\ve(\Ga)}\mu_a
\e{22}
where $\mu_a$ is a constant infinitesimal fermionic parameter.
(One has to make use
of \r{14} and \r{16}.)

In the case when $\mu_a$ depends on $\Ga$ and $\la$ the
transformation \r{22} induces a
change in  the gauge fixing action $X$  after the additional transformation
\be
&&\del_1\Ga=\halv(\Ga, \del F_a)^a,\;\;\;\del F_a(\Ga)\equiv
\frac{2\hbar}{i}\mu_a(\Ga, \la) \e{240}
is performed. This change is given by
\be
&&\del X=(X, \del F_a)^a-V^a\del F_a-i\hbar\Delta^a\del F_a
\e{241}
One may now show that
\be
&&(X, \del X)^a-V^a\del X=i\hbar\Delta^a \del X
\e{23}
provided $\del F_a$ is chosen to have the following form
\be
&&\del  F_a=\ve_{ab}\left\{(X, \del\Xi)^b-V^b\del\Xi
-i\hbar\Delta^b\del\Xi\right\}
\e{242}
The gauge independence of the functional integral is then  confirmed.

In order to illustrate the above properties we consider a particular
solution of \r{16}.
In fact, a simple natural solution is given by
\be
X=G_A\la^A+{\cal K}Y
\e{191}
where
\be
&&Y=\bar{\phi}_A\phi^A-2F(\phi)
\e{20}
\be
&&G_A=\bar{\phi}_A-F(\phi)\frac{\stackrel{\lea}{\partial}}{\partial\phi^A}
\e{21}
where $F(\phi)$ is an arbitrary function of the original fields $\phi^A$.
In this case the transformation \r{22} is given by
\be
&&\del\phi^A=-\left(\frac{\stackrel{\ra}{\partial}W}{\partial\phi^*_{Aa}}-
\frac{3}{2}\pi^{Aa}\right)\mu_a
\e{25}
\be
&&\del\phi^*_{Aa}=\mu_a\left(\frac{\stackrel{\ra}{\partial}W}
{\partial\phi^A}+F\frac{\stackrel{\lea}{\partial}}
{\partial\phi^A}\frac{\stackrel{\lea}{\partial}}{\partial\phi^B}\la^B\right)
\e{26}
\be
&&\del\pi^{Aa}=\ve^{ab}\left(\frac{\stackrel{\ra}{\partial}W}
{\partial\bar{\phi}_A}-\la^A\right)\mu_b
\e{27}
\be
&&\del\bar{\phi}_A=\mu_a\left(\ve^{ab}\frac{\stackrel{\ra}{\partial}W}
{\partial\pi^{Ab}}
+\halv\ve^{ab}\phi^*_{Ab}-\pi^{Ba}F
\frac{\stackrel{\lea}{\partial}}{\partial\phi^B}
\frac{\stackrel{\lea}{\partial}}{\partial\phi^A}\right)
\e{28}
It is now straight-forward to check that the integrand of \r{15}
is invariant under these
transformations when $\mu_a$ is constant. However,
if we now let $\mu_a$ depend on
$\phi^A$ and $\pi^{Aa}$, then
the transformations \r{25}-\r{28} lead to the following  change
of the integrand in \r{15}
\be
&&\mu_a\frac{\stackrel{\lea}{\partial}}{\partial\phi^A}
\left(\frac{\stackrel{\ra}{\partial}W}{\partial\phi^*_{Aa}}-
\frac{3}{2}\pi^{Aa}\right)+
\ve^{ab}\mu_a\frac{\stackrel{\lea}{\partial}}{\partial\pi^{Aa}}
\left(\frac{\stackrel{\ra}{\partial}W}{\partial\bar{\phi}_A}-\la^A\right)
\e{29}
On the other hand, by means of the additional transformations \r{240}
here given by
\be
&&\del_1\phi^*_{Aa}=
-\frac{\hbar}{i}\mu_a\frac{\stackrel{\lea}{\partial}}{\partial\phi^A},
\;\;\;\del_1\bar{\phi}_{A}=
-\frac{\hbar}{i}\ve^{ab}\mu_a
\frac{\stackrel{\lea}{\partial}}{\partial\pi^{Ab}}
\e{30}
eq.\r{29} acquires the form
\be
&&-2\mu_a\frac{\stackrel{\lea}{\partial}}{\partial\phi^A}
\pi^{Aa}-2\ve^{ab}\mu_a
\frac{\stackrel{\lea}{\partial}}{\partial\pi^{Ab}}   \la^A
\e{31}
Finally choosing the ansatz
\be
&&\mu_a=\frac{i}{2\hbar}\del F_a(\phi,\pi),\;\;\;\del F_a(\phi,\pi)=
-\frac{1}{2}\ve_{ab}\del
F\frac{\stackrel{\lea}{\partial}}{\partial\phi^B}  \pi^{Bb}
 \e{32}
which corresponds to the choice $\del\Xi=-\halv\del F$ in \r{242},
one may show that \r{31} corresponds to the variation
\be
&&F\:\ra\:F+\del F
\e{33}
in the integrand of the functional integral \r{15}. Thus, we have
proved that the functional integral
\r{15} is gauge independent. One may notice that we in our proof
of gauge independence only have
made use of superpositions of purely anticanonical transformations
and the ones generated by the
operator $V^a$ in \r{8}.

        \section{Conclusion.}
Our main results are as follows: First, we have
identified the former "parametric"
variables $\pi^{Aa}$ with fields anticanonically conjugate to the antifields
$\bar{\phi}_A$ in the usual sense. The antibrackets $(F,G)^a$ as well as
the operators
$\triangle^a$, $V^a$ are then extended to
include these new anticanonical pairs. The
resulting new $V^a$ turns out to act as a derivative on the new antibracket and
to anticommute with $\Delta^a$ without symmetrization (eq:s (9) and (11)).
We cast then
the gauge fixing mechanism  into a completely anticanonical form, and give the
conditions under which the functional integral is gauge independent. Finally,
we
verify the formalism for a simple set of allowed hypergauge  conditions. One
may
notice that although we have tripled the number of field variables, the number
of
hypergauge  functions multiplying the Lagrange multipliers are only those of
the
original fields.

The formulation presented in this paper covers the case when the hypergauge
conditons satisfy an abelian algebra in terms of the antibrackets, and when
the field-antifield phase space is spanned by
anticanonical pairs. The generalization
to a geometric description in terms of general
 phase spaces as well as to arbitrary
nonabelian hypergauge conditions is given in \cite{BMS}.\\

\vspace{5mm}

 {\bf Acknowledgement:}  I.B. would like to thank Prof. Lars Brink for kind
hospitality at the
Institute of Theoretical Physics, Chalmers University, G\"{o}teborg.
I.B. was supported in part by the Board of Trustees of ICAST and Nato
Linkage Grant \#931717.

            \newpage
    \noindent
    {\Large{\bf{Appendix}}}\\
     \vspace{5mm}
{\bf Identities for the proof of eq.(26)}

In order to prove that the quantum master
eq.\r{16} is invariant under the variation
\r{241} with $\del F_a$ given by \r{242} one needs the following identities
\be
&&\ve_{bc}(X, (X, (X, \del\Xi)^c)^b)^a\equiv\ve_{bc}\left\{(X, (\del\Xi, (X,
X)^b)^c)^a+\right.\nn\\
&&\left.+(\del\Xi, (X, (X, X)^b)^c)^a+2((X, X)^b, (X, \del\Xi)^c)^a\right\}
\e{34}
\be
&&\ve_{bc}\left\{\Delta^a(X, (X, \del\Xi)^c)^b+(\del\Xi, (X, \Delta^c
X)^b)^a+(\Delta^c X, (X, \del\Xi)^b)^a-(X, (\Delta^cX,
\del\Xi)^b)^a+\right.\nn\\
&&\left.+(X, (X, \Delta^c\del\Xi)^b)^a+(X,
\Delta^b(X, \del\Xi)^c)^a-(\Delta^b X, (X, \del\Xi)^c)^a         \right\}
\equiv\nn\\
&&\equiv\ve_{bc}\halv\left\{\Delta^a(\del\Xi, (X, X)^b)^c+(\del\Xi, \Delta^c(X,
X)^b)^a+2(\Delta^b\del\Xi, (X, X)^c)^a
\right\}
\e{35}
\be
&&\ve_{bc}\left\{\Delta^a\Delta^b(X, \del\Xi)^c-
\Delta^a(\Delta^cX, \del\Xi)^b+2(\Delta^cX,
\Delta^b\del\Xi)^a+\right.\nn\\
&&\left.+(\del\Xi, \Delta^b\Delta^cX)^a+\Delta^a(X,
\Delta^c\del\Xi)^b+(X, \Delta^b\Delta^c\del\Xi)^a    \right\}
\equiv 0
 \e{36}
They may be derived by direct use of the definition \r{1} of the antibracket,
or
indirectly from \r{6} by means of the identities
$\Delta^a\Delta^b\Delta^c(X^3\del\Xi)\equiv 0$, $\Delta^a\Delta^b
\Delta^c(X^2\del\Xi)\equiv 0$
and $\Delta^a\Delta^b\Delta^c(X\del\Xi)\equiv 0$
respectively. One may notice that
these identities are even valid when $\Delta^a$
is replaced by $\Delta^a+\al V^a$ for
any constant $\al$.


\begin{thebibliography}{Fierz}

\bibitem{BLT1}I. A. Batalin, P. M. Lavrov and I. V. Tyutin, \ {\sl J. Math.
            Phys.}\ {\bf 31}, 1487 (1990)


            \bibitem{BLT2}I. A. Batalin, P. M. Lavrov and I. V. Tyutin,
\ {\sl J. Math.
            Phys.}\ {\bf 32}, 532 (1991)

\bibitem{BLT3}I. A. Batalin, P. M. Lavrov and I. V. Tyutin, \ {\sl J. Math.
            Phys.}\ {\bf 32}, 2513 (1991)


     \bibitem{MH}M. Henneaux,  \  {\sl Phys. Lett.}\ {\bf B282},\
         372\ (1992)



\bibitem{BCG}G. Barnich, R. Constantinescu and P. Gr\'{e}goire, \ {\sl
 Phys. Lett.}\ {\bf
           B293}, \ 353 (1992)



\bibitem{GH}P. Gr\'{e}goire and M. Henneaux, \ {\sl J.  Phys.}\ {\bf
           A26}, \ 6073 (1993)

 \bibitem{DD}P. H. Damgaard and F. De Jonghe, \ {\sl Phys. Lett.}\ {\bf
            B305}, \ 59 (1993)



     \bibitem{LT}L. T\u{a}taru,
             \ {\em A cohomological approach to the Batalin-Lavrov-Tyutin
covariant quantization of
irreducible gauge theory.} Preprint TUW 94-18


       \bibitem{BMS}I. A. Batalin, R. Marnelius and A. M. Semikhatov, \ {\em
Triplectic quantization:  A geometrically
covariant description of Sp(2)-symmetric
Lagrangian formalism.} \ ITP-G\"{o}teborg preprint 94-31. Hep-th/9502031







            \end{thebibliography}
    \end{document}